\documentstyle[twoside,fleqn,espcrc2]{article}

\input BoxedEPS
\SetRokickiEPSFSpecial  
\HideDisplacementBoxes

\makeatletter
\long\def\@makefigurecaption#1#2{\vskip 6mm #1. #2\par}

\def\fnum@figure{Figure \thefigure}
\def\figure{\let\@makecaption\@makefigurecaption \@float{figure}}
\@namedef{figure*}{\let\@makecaption\@makefigurecaption \@dblfloat{figure}}

\textfloatsep=\floatsep           
\intextsep=\floatsep              

\makeatother

\title{Evidence of Instanton Effects in Hadrons from the\protect\linebreak
 Study of Low
 Eigenfunctions of the Dirac Operator}

\author{T.L. Ivanenko 
        and 
        J.W. Negele\address{Center for Theoretical Physics\\ 
Laboratory for Nuclear Science
and Department of Physics\\ 
Massachusetts Institute of Technology, Cambridge
MA 02139\hfill MIT-CTP\#2677\quad  hep-lat/9709130}%
\thanks{This work is
supported in part by funds provided by the US Department of Energy (DOE) 
under cooperative research agreement \#DF-FC02-94ER40818.}}
       
\begin{document}

\begin{abstract}
To elucidate the role in hadron structure of quark zero modes associated with
instantons, the lowest eigenfunctions of the Dirac operator have been calculated in
quenched QCD, full QCD, and full QCD with cooling. Eigenmodes associated with a
single instanton and an instanton--anti-instanton pair were studied to understand
the role of lattice artifacts for Wilson Fermions. By truncating the spectral
representation of the  quark propagator, we show that the rho and pion
contributions to the vector and pseudoscalar correlation functions are dominated
by the zero-mode contributions.
\end{abstract}

\maketitle

\section{Introduction and Motivation}

This work is part of an ongoing effort to use the lattice to obtain insight into
the structure of light  hadrons by finding the paths that dominate the QCD
path integral\cite{R:JN:12}. The picture that emerges is that of the instanton
liquid developed by Shuryak\cite{R:JN:03} and by Dyakanov and
Petrov\cite{R:JN:05}. In this picture, light quarks propagate by hopping
between zero modes, and the resulting 't~Hooft interaction\cite{R:JN:08}
accounts for many phenomena, ranging from the channel dependence of
vacuum hadron current correlation functions\cite{R:JN:04} to the so-called
``spin crisis" in which helicity is transferred from valence quarks to 
gluons and $\bar q q$ excitations and the  $\bar d d$ to $\bar u u$ ratio in
the proton. Previous calculations show that removing non-instanton
components of  the gluon configurations by cooling\cite{R:JN:11} reproduces
the essential features of the instanton model and observables calculated
with  all gluons are similar to those obtained using only instantons that
survive cooling\cite{R:JN:07}.   

The chief limitation of these cooling studies is
that although the loss of a single instanton by cooling can be avoided by the
use of an improved action\cite{R:JN:14}, the annihilation of
instanton--anti-instanton pairs is unavoidable\cite{R:JN:17}.
Hence, the purpose of this work is to avoid the limitations of cooling by
studying the low eigenfunctions of the Dirac operator directly in uncooled
configurations. 

One essential piece of background information concerns vacuum matrix
elements of meson  correlation  functions $\langle \Omega
| T J(x) J(0)^{\dagger}| \Omega \rangle$ in the vector and pseudoscalar
channels\cite{R:JN:07}. Lattice calculations agree well with
phenomenological dispersion analyses\cite{R:JN:03}, and the strength in the
vicinity of 1~fm is strongly dominated by the rho and pion 
contributions. 

\section{Eigenmodes of the Dirac operator}

Using the  k-step Arnoldi method\cite{R:JN:16} we have calculated eigenmodes of
the Wilson Dirac operator 
\begin{eqnarray*}
 D \psi_x &=&
 \psi_x - \kappa \sum_\mu \Bigl[ (r-\gamma_\mu) u_{x,\mu}
\psi_{x+\mu} \\
&& {} + (r + \gamma_\mu) 
u^\dagger_{x-\mu,\mu} \psi_{x-\mu} \Bigr]
\end{eqnarray*}
in several simple background configurations to understand their structure,
as well as on $16^4$ lattices for quenched QCD at $\beta=5.85$ and for
unquenched QCD at $\beta=5.5$ and $\kappa_{sea}=0.16$.  

\begin{figure}[htb]
\BoxedEPSF{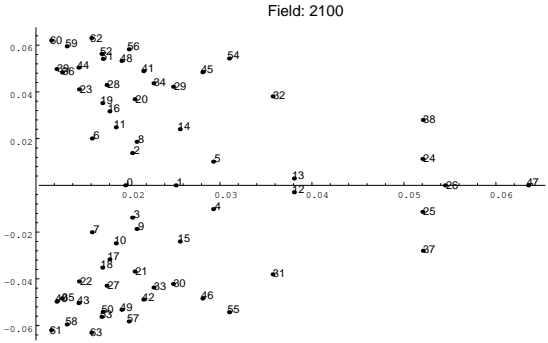}
\caption{Lowest 64 eigenvalues of the Dirac operator for an
unquenched gluon configuration.}
\label{F:TIJN:1}
\end{figure}

\begin{figure}[htb]
\BoxedEPSF{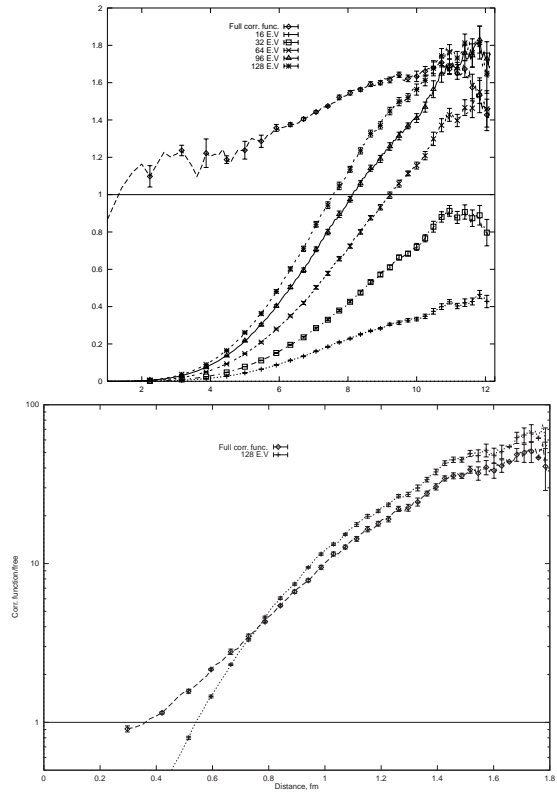}
\caption{Contributions of low Dirac eigenmodes to the vector (upper graph)
  and pseudoscalar (lower graph) vacuum correlation functions.}
\label{F:TIJN:2}
\end{figure}

As is well known, in the
free case, a discrete set of points starting at the lowest Matsubara mode
$(0,0,0,\pm{\pi/ L_t})$ approximates  the continuum spectrum 
$\frac{1}{m}[m+i |\vec{p}|]$  in the physical regime and 
unphysical fermion modes are pushed to large (real) masses.   In the
presence of an instanton of size $\rho$ at $x=0$, there are two lattice
artifacts. Because of the Wilson term, the eigenvalue acquires a shift of
approximately $Re \lambda_0 = \langle \psi_0|D|\psi_0\rangle ={{3 \kappa
a} / {\rho^2}}$. Also, in addition to a mode with zero
imaginary part that approaches the continuum result 
 $$ 
 \psi_0(x)_{s,\alpha} = u_{s, \alpha} \frac{\sqrt{2}}{\pi} \frac{\rho}{(x^2 +
 \rho^2)^{3/2}}
 $$
one spin and color component of the lowest Matsubara mode  $(0,0,0,
\pm{\pi/ L_t})$ mixes with the zero mode with a coefficient that goes
to zero in the large volume limit.  The eigenvalues for an
instanton--anti-instanton pair are approximately
 $$ \lambda = {{\lambda_I+\lambda_A}\over 2}
 \pm\sqrt{{{(\lambda_I-\lambda_A)^2} \over 4} -(T_{IA})^2}
 $$
so that when the interaction $T_{IA}$ exceeds the difference in the Wilson
term shifts, complex conjugate pairs of eigenvalues move slightly off the
real axis.  

Fig.~\ref{F:TIJN:1} shows the lowest 64 complex eigenvalues of the Dirac operator on
an unquenched gluon configuration. Note that the eigenvalues of this uncooled
configuration have just the structure expected from the pure one- and  two-instanton
studies above: isolated zeros along the real axis corresponding to  individual instantons
and complex conjugate pairs corresponding to instanton--anti-instanton pairs.    To set
the scale, note that the lowest Matsubara
mode would occur at 0.06 on the imaginary axis, corresponding to 380 MeV, so all
the conjugate pairs below this value are understood to be the results of zero modes.

\section{Zero mode expansion}
The Wilson--Dirac operator has the property that $D=\gamma_5 D^\dagger
\gamma_5$, which implies that $\langle \psi_j |\gamma_5| \psi_i \rangle=0$ unless
$\lambda_i = \lambda^*_j$ and we may write the spectral representation of the
propagator
$$
\langle x |D^{-1}|y \rangle = \sum_i \frac{\langle x|\psi_i \rangle \langle \psi_{\bar{i}}
|\gamma_5| y\rangle}{\langle \psi_{\bar{i}}|\gamma_5| \psi_i\rangle \lambda_i}
$$
where $\lambda_i = \lambda_{\bar{i}}^*$.  A clear indication of the role of zero modes
in light hadron observables is the degree to which truncation of the expansion to the
zero mode zone reproduces the result with the complete propagator.

Fig.~\ref{F:TIJN:2} shows the result of truncating the vacuum correlation
functions for the vector and pseudoscalar channels to include only low
eigenmodes\cite{R:JN:15}. On a $16^4$ lattice, the full propagator contains
786,432 modes. The top plot of Fig.~\ref{F:TIJN:2}  shows the result of including
the lowest 16, 32, 64, 96, and finally 128 modes for an unquenched
configuration with a 63~MeV valence quark mass. Note that the first 64 modes
reproduce most of the strength at 1~fm (10 lattice spacings) where the
correlator is saturated by the  rho resonance peak, and by the time we include
the first 128 modes, all the strength is accounted for.  Similarly, the lower plot
for a quenched configuration with a 23~MeV  quark mass shows that
the lowest 128 modes also account for  the analogous pion contribution to the
pseudoscalar vacuum correlation function.  Thus, without having to resort to
cooling, by looking directly at the contribution of the lowest eigenfunctions, we
have shown that the zero modes associated with instantons dominate the
propagation of rho and pi mesons in the QCD vacuum.

\subsection{Localization}

Finally, it is interesting to ask whether the lattice zero mode eigenfunctions are
localized on instantons.  
  This was studied by plotting the quark density distribution for
  individual eigenmodes in
  the $x$-$z$ plane for all values of $y$ and~$t$, and comparing with
  analogous plots of the action density. As expected, for a cooled
  configuration the eigenmodes 
correspond to
linear combinations of localized zero modes at each of the instantons.
 What is truly remarkable, however, is
 that the eigenfunctions of the uncooled configurations also
exhibit localized peaks at locations at which instantons are identified by
cooling.  Thus, in spite of the fluctuations several orders of magnitude larger
than the instanton fields themselves, the light quarks essentially average
out these fluctuations and produce localized peaks at the topological
excitations.  When one analyzes a number of eigenfunctions, one finds that
all the instantons remaining after cooling correspond to localized quark
fermion peaks in some eigenfunctions.  However, some fermion peaks are
present for the initial gluon configurations that do not coincide with
instantons that survive cooling, corresponding to
instanton--anti-instanton pairs that were annihilated during cooling.

\subsubsection*{Acknowledgments}
It is a pleasure to acknowledge  the donation by
Sun Microsystems of the 24 Gflops E5000 SMP cluster on which the most recent
calculations were performed and the computer resources provided by NERSC with
which this work was begun.

\end{document}